\documentclass[11pt,letter]{emulateapj}
\usepackage{amsmath}
\usepackage{graphicx}
\usepackage{epstopdf}

\shorttitle{Polarization Gradients in the Galactic Plane}
\shortauthors{Herron et al.}
\begin{document}

\title{Polarization Gradient Study of Interstellar Medium Turbulence Using The
Canadian Galactic Plane Survey}

\author{C.~A.~Herron\altaffilmark{1}, J.~Geisbuesch\altaffilmark{2,3}, T.~L.~Landecker\altaffilmark{2,1},
R.~Kothes\altaffilmark{2}, B.~M.~Gaensler\altaffilmark{4,1}, G.~F.~Lewis\altaffilmark{1}, N.~M.~McClure-Griffiths\altaffilmark{5} \& E.~Petroff\altaffilmark{6,1}}

\altaffiltext{1}{Sydney Institute for Astronomy, School of Physics, A28, The University
of Sydney, NSW, 2006, Australia; C.Herron@physics.usyd.edu.au}

\altaffiltext{2}{National Research Council Canada, Herzberg Astronomy \&
Astrophysics, Dominion Radio Astrophysical Observatory, P.O. Box 248, Penticton,
British Columbia, V2A 6J9, Canada} 

\altaffiltext{3}{Present address: Karlsruhe Institute for Technology, PO Box 3640, 76021 Karlsruhe, Germany}

\altaffiltext{4}{Dunlap Institute for Astronomy and Astrophysics, University of Toronto, 50 St. George Street,
Toronto, Ontario, M5S 3H4, Canada}

\altaffiltext{5}{Research School of Astronomy and Astrophysics, The Australian National University, Canberra, ACT 2611, Australia}

\altaffiltext{6}{ASTRON Netherlands Institute for Radio Astronomy, Postbus 2, NL-7990 AA Dwingeloo, The Netherlands}

\date{\today}



\begin{abstract}
We have investigated the magneto-ionic turbulence in the interstellar medium through spatial gradients of the complex radio polarization vector in the Canadian Galactic Plane Survey (CGPS). The CGPS data cover 1300 square-degrees, over the range ${53^{\circ}}\leq{\ell}\leq{192^{\circ}}$, ${-3^{\circ}}\leq{b}\leq{5^{\circ}}$ with an extension to ${b}={17.5^{\circ}}$ in the range ${101^{\circ}}\leq{\ell}\leq{116^{\circ}}$, and arcminute resolution at 1420~MHz. Previous studies found a correlation between the skewness and kurtosis of the polarization gradient and the Mach number of the turbulence, or assumed this correlation to deduce the Mach number of an observed turbulent region. We present polarization gradient images of the entire CGPS dataset, and analyze the dependence of these images on angular resolution. The polarization gradients are filamentary, and the length of these filaments is largest towards the Galactic anti-center, and smallest towards the inner Galaxy. This may imply that small-scale turbulence is stronger in the inner Galaxy, or that we observe more distant features at low Galactic longitudes. For every resolution studied, the skewness of the polarization gradient is influenced by the edges of bright polarization gradient regions, which are not related to the turbulence revealed by the polarization gradients. We also find that the skewness of the polarization gradient is sensitive to the size of the box used to calculate the skewness, but insensitive to Galactic longitude, implying that the skewness only probes the number and magnitude of the inhomogeneities within the box. We conclude that the skewness and kurtosis of the polarization gradient are not ideal statistics for probing natural magneto-ionic turbulence.
\end{abstract}

\section{Introduction} 
\label{intro} 

The Galactic interstellar medium (ISM) is a complex multi-phase environment with (at least) molecular, cold neutral, warm neutral, warm ionized and hot ionized phases, and the whole is threaded by a magnetic field that is itself a significant energy-carrying constituent (\citealp{Ferriere2001}, \citealp{Heiles2012}). The phases are defined by their temperatures, densities and pressures, varying over orders of magnitude. Although globally in equilibrium, there are strong local departures from this state, and the whole medium is turbulent over a large range of physical scales \citep{Cox2005}. Turbulence is an effective distributor of energy between different physical scales, and its study is important for our detailed and global understanding of the ISM and the processes within it.

The Canadian Galactic Plane Survey (CGPS, \citealt{Taylor2003}) has provided data on the major constituents of the ISM, bringing together radio, millimetre, and infrared surveys at arcminute resolution. In this paper we use the data on polarized radio continuum emission at 1420~MHz, observed with the Synthesis Telescope (ST) at the Dominion Radio Astrophysical Observatory (DRAO), to study the turbulent structure of the ISM on a large range of spatial scales. The CGPS polarization images \citep{Landecker2010}, and other polarization data at decimetre wavelengths (\citealp{Gaensler2011}, \citealp{Iacobelli2014}, \citealp{Sun2014}), show widespread small-scale structure in and near the Galactic plane that has no counterpart in total intensity. The common interpretation is that these structures arise from Faraday rotation, the rotation of the plane of polarization as light propagates parallel to a magnetic field in an ionized medium, which is characterized by the rotation measure (RM). These small-scale structures are believed to reflect the turbulent state of the magneto-ionic medium. 

Statistical techniques have been used to investigate turbulence within the magneto-ionic medium. These methods have used observations of the Faraday rotation of background sources as well as data on the extended polarized emission, and have used correlation or structure-function methods of analysis (e.g.~\citealp{Baccigalupi2001}, \citealp{Haverkorn2004}, \citealp{Haverkorn2006}, \citealp{Stil2011}, \citealp{Carretti2011}, and \citealp{Stutz2014}).

In this paper we apply a statistical tool which is complementary to other methods, the polarization gradient method, developed by \citet{Gaensler2011}. In this technique the gradient of the Stokes vector $(Q, U)$ is calculated, and this provides a unique view of turbulence in the magnetized and ionized interstellar gas. The magnitude of the gradient remains unaffected by rotation or translation in the $(Q, U)$ plane, which can be caused by superimposed foreground emission or Faraday rotation, and so can reveal polarization characteristics that would otherwise be obscured. \citet{Burkhart2012} extended this work by analyzing synthetic data from MHD simulations, and demonstrated that the gradient technique provides an indicator of the regime of turbulence in the ISM (the behaviour of the plasma motions as defined by the sonic and Alfv\'enic Mach numbers). In particular, \cite{Burkhart2012} found that the skewness and kurtosis of the gradient data can provide a measure of the Mach number, ${\cal M}_s$. 

Our work, examining the polarization gradients of a large area of the Galactic plane in the outer Galaxy, joins three other studies that have looked at substantial areas. \citet{Sun2014} applied the gradients technique to $240$ square degrees of the Galactic plane at low longitudes, examining two datasets at $2300$ and $4800$~MHz, both with a resolution of about $10'$. \citet{Iacobelli2014} analyzed the S-PASS dataset, a $2300$~MHz survey of polarized emission from the entire sky south of declination $0^{\circ}$, $2\pi$~steradians, with an angular resolution of $10.8'$. The polarization gradients in the CGPS data were previously discussed by \cite{Robitaille2015}, who analyzed a $56$ square degree portion of the CGPS using wavelet analysis. They found that the network of polarization gradient filaments in their portion of the CGPS is sensitive to angular resolution.

In this paper we use the polarization gradient method to analyze the CGPS data over an area of $1300$ square degrees near the Galactic plane, at an angular resolution of $\sim{150''}$, surpassing that of earlier studies. We show that the skewness and kurtosis of the polarization gradient are not reliable probes of observed magneto-ionic turbulence, due to their sensitivity to inhomogeneities.

The paper is organized as follows. Section~\ref{data} describes relevant details of data acquisition and processing, and Section~\ref{method} reviews the polarization gradient method and introduces the statistical analysis used to derive information on ISM turbulence from the data. We demonstrate the effects of angular resolution in Section~\ref{resstudy}, and present the polarization gradient for the entire CGPS in Section~\ref{allsky}. In Section \ref{statgrad} we present the method and results of our statistical analysis of the polarization gradient maps, and discuss our findings in Section~\ref{discuss}.

\section{Observations and data processing}
\label{data}

The DRAO ST acquires polarization data from baselines between 13 and 617~m, and consequently has excellent surface-brightness sensitivity to structures of sizes $1'$ to ${\sim}40'$. Information on larger structures has been incorporated from observations with the Effelsberg 100-m \citep{Reich2004} and the DRAO 26-m \citep{Wolleben2006} telescopes. The methods used in the data combination and the steps taken to ensure accurate calibration of the three intensity scales are described in detail by \citet{Landecker2010}. While that paper deals with the longitude range ${65^{\circ}}\leq{\ell}\leq{175^{\circ}}$, the same procedures were used for the entire longitude range, ${53^{\circ}}\leq{\ell}\leq{192^{\circ}}$, discussed here. The amplitude scales of the three contributing data sets (DRAO ST, Effelsberg, and DRAO 26-m) are considered matched within 10\%. Key properties of the CGPS are provided in Table \ref{tab:surprop}.\footnote{The CGPS data are available at the Canadian Astronomy Data Centre: http://www.cadc-ccda.hia-iha.nrc-cnrc.gc.ca/en/cgps/}

\begin{table}
\caption[]{Survey properties of the CGPS relevant to this work.}
\label{specs}
\begin{center}
\begin{tabular}{ll}
\hline
Coverage                     & ${53^{\circ}} < {\ell} < {192^{\circ}},
                               {-3^{\circ}} < {b} < {5^{\circ}}$ \\
                             & ${101^{\circ}} < {\ell} < {116^{\circ}},
                               {5.0^{\circ}} < {b} < {17.5^{\circ}}$ \\
Continuum bandwidth          & 30\,MHz in four bands of \\
                             & 7.5\,MHz each \\
Polarization products        & Stokes $I$, $Q$, and $U$ \\
Center frequencies           & 1407.2, 1414.1, 1427.7,\\
                             & and 1434.6\,MHz \\
Angular resolution           & $58'' \times 58''$\,cosec\thinspace$\delta$ \\
Sensitivity, $I$             & 200 to 400 $\mu$Jy/beam rms \\
Sensitivity, $Q$ and $U$     & 180 to 260 $\mu$Jy/beam rms \\
Typical noise in mosaicked   & \\
images                       &  ${76}\thinspace{\rm{sin{\thinspace}(declination)}}$~mK \\
Sources of single-antenna data & Effelsberg 100-m Telescope \\
                               & and DRAO 26-m Telescope \\
\hline
\end{tabular}
\end{center}
\label{tab:surprop}
\end{table}

In places where the total intensity was high, the CGPS polarization data were reprocessed, using the algorithm described by \cite{Reid2008}, to improve the correction for leakage of energy from Stokes $I$ to Stokes $Q$ and $U$. Other processing uses the DRAO Export Package (\citealp{Higgs1997}) and tailor-made routines.

The CGPS was observed on a hexagonal grid of antenna pointings in Galactic co-ordinates, spaced $112'$ apart \citep{Taylor2003}. Consequently, the sensitivity of the mosaicked data varies by $\pm{20\%}$ over the survey area. In regions of low polarized intensity, or low intensity of the spatial gradients of polarization, a hexagonal pattern of noise is sometimes evident, and this pattern can be enhanced by the differentiation inherent in the calculation of spatial gradients (see equation~\ref{eqno:nablap}).

The angular resolution of the DRAO ST changes with cosec\thinspace$\delta$ and the sensitivity to small structures in polarization gradients changes with declination and with orientation of features relative to the elliptical beam. 

The gradient method cannot recover structure lost to depolarization, which occurs due to the vector averaging of polarized signals with different polarization angles within the telescope beam (referred to as beam depolarization), within the passband, or within the emitting source. For the DRAO ST, a rotation measure (RM) value of 300 rad m$^{-2}$ produces bandwidth depolarization of 8\% when data from all four bands are averaged, and less than 0.3\% in a single band. The Galactic RM is less than this over the entire area discussed in this paper (\citealp{Brown2003}, \citealp{Oppermann2012}). Beam depolarization is unavoidable, and will become more severe as the angular resolution degrades, but the angular resolution of $1'$ will reveal parsec-scale structure at a distance of several kiloparsecs, adequate for the nearby Galaxy. 

\section{The polarization gradient method}
\label{method}

The complex polarization vector is 
\begin{equation}
 {\bf{P}} = Q+iU = |{\bf{P}}| e^{2i\theta},
\label{equ:cmplxP}
\end{equation}
where $|{\bf{P}}|$ is the vector amplitude, and $\theta$ is the observed polarization angle of the radiation.

The polarization gradient is defined as the modulus of the gradient of the complex polarization vector \citep{Gaensler2011}. It is
\begin{equation}
{|\nabla {\bf P}|}={\sqrt{\biggl(\frac{\partial Q}{\partial x}\biggr)^2
+\biggl(\frac{\partial U}{\partial x}\biggr)^2
+\biggl(\frac{\partial Q}{\partial y}\biggr)^2
+\biggl(\frac{\partial U}{\partial y}\biggr)^2}.}
\label{eqno:nablap}
\end{equation}

\cite{Burkhart2012} calculated the skewness and kurtosis of the probability density function (PDF) of $|\nabla {\bf{P}}|$, where the skewness of a PDF constructed from $N$ values $X_i$ is given by
\begin{equation}
{\gamma}={{\frac{1}{N}}{\thinspace}{\sum^{N}_{i=1}\left(\frac{X_i-\mu}{\sigma}\right)^3}},
\label{equ:skew}
\end{equation}
and the excess kurtosis is 
\begin{equation}
{\beta}={{\frac{1}{N}}{\thinspace}{\sum^{N}_{i=1}\left(\frac{X_i-\mu}{\sigma}\right)^4} -
3.}
\label{equ:kurt}
\end{equation}
Here $\mu$ and $\sigma$ are the mean and standard deviation (1st and 2nd moment) of the distribution:
\begin{eqnarray}
\mu & = & {\frac{1}{N}}{\thinspace}\sum^{N}_{i=1}X_i \\
\sigma & = & \sqrt{{\frac{1}{N-1}}{\thinspace}\sum^{N}_{i=1}\left(X_i-\mu\right)^2}.
\end{eqnarray}

Based on MHD simulations, \citet{Burkhart2012} found that the skewness and kurtosis of $|\nabla {\bf{P}}|$ are generally higher for larger sonic Mach number ${\cal M}_s \equiv \langle |{\bf v}|/c_s \rangle$, where ${\bf v}$ is the local velocity, $c_s$ is the sound speed, and the averaging is done over the entire region of interest (see their Figure 7). \citet{Burkhart2012} also show that high angular resolution is important to unambiguously distinguish different regimes of turbulence, and demonstrate that a ``double-jump'' structure in the polarization gradient (their Figure 4) is a characteristic of a strong shock and supersonic turbulence.

\section{Angular Resolution Dependence of Polarization Gradients}
\label{resstudy}

\begin{figure*}
\begin{center}
\includegraphics[scale=0.8]{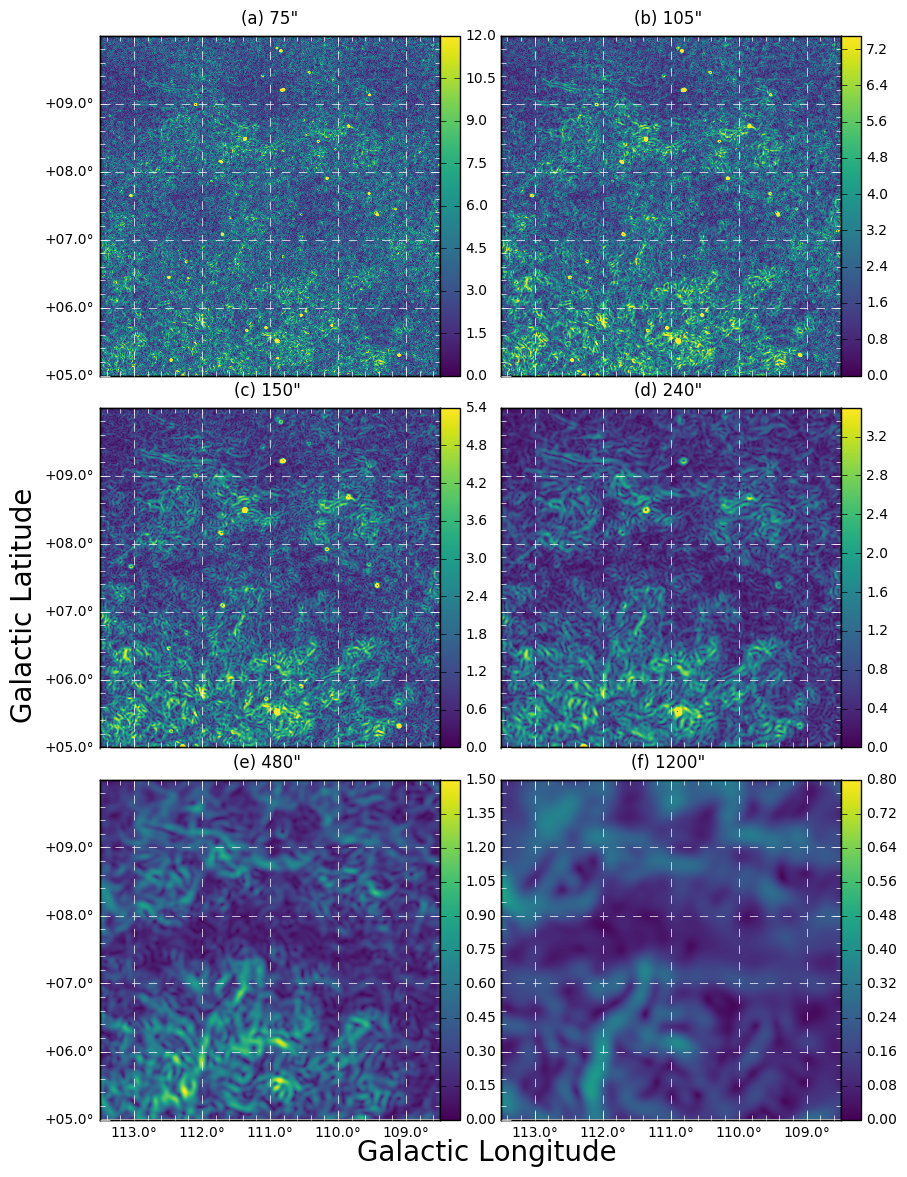}
\caption{Polarization gradient images of a section of the northern latitude extension of the CGPS, at various angular resolutions. The angular resolutions are a) 75'', b) 105'', c) 150'', d) 240'', e) 480'', f) 1200''. The colorbars give the magnitude of the polarization gradient, expressed in K/degree.}
\label{pgradres}
\end{center}
\end{figure*}

Smoothing the data to a broader angular resolution has two desirable effects: it reduces noise in the gradient images and decreases the impact of the non-uniformity of the noise on those images. However, as mentioned in Section \ref{data}, we also need to maintain as high a resolution as possible, so that we are able to distinguish different regimes of turbulence. To find the best compromise between these objectives, we smoothed the $Q$ and $U$ data for the CGPS to various resolutions by convolving with a circular Gaussian, up to a maximum final resolution of $20'$. We then produced images of the polarization gradients at each resolution using equation \ref{eqno:nablap}. Figure~\ref{pgradres} displays polarization gradient maps derived for the same CGPS region at different resolutions, in units of K/degree.

Near the original resolution of the data, we find that very few polarization gradient structures can be seen, and instead we see the hexagonal grid of antenna pointings (above $b = 7^{\circ}$ in Figure \ref{pgradres}a). This is because the process of differentiation amplifies the noise between the pointings, so that the polarization gradient filaments are difficult to observe. As the angular resolution worsens, the noise is suppressed, and polarization gradient structures become clearer, as seen at angular resolutions of $105''$ and $150''$. Further increases in the smoothing scale cause more polarization gradient structures to become apparent in previously noisy regions, and previously visible polarization gradient structures grow in size, until they begin to overlap. At the poorest angular resolutions studied, the polarization gradient structures do not resemble the structures seen at the best angular resolutions (see \citealt{Robitaille2015} for additional discussion). 

We also see that the peak amplitude of $|\nabla{\bf{P}}|$ in these images decreases with increasing smoothing scales from the original resolution of the data to $20'$. This is expected, as smoothing the maps of $Q$ and $U$ reduces the amplitude of fluctuations in these quantities, and hence reduces the derivatives that are used to calculate the polarization gradient. Polarized point sources are also visible in the high angular resolution polarization gradient maps, but the smoothing applied to create the poor angular resolution maps means that point sources are not visible in these maps.

By studying the polarization gradient maps produced at various resolutions, we decided that a resolution of $150''$ was ideal. At this angular resolution, most of the noisy regions in the CGPS have been suppressed, so that true polarization gradient structures are visible. Increasing the smoothing scale to $210''$ does cause more polarization gradient filaments to become visible in the remaining noisy regions, however it also causes previously visible, small-scale structures to become distorted. Hence, we believe that $150''$ is the optimal angular resolution for studying the polarization gradient filaments visible in the CGPS data.

We note that as the beam is elliptical at the highest and lowest longitudes of the survey, the actual resolution varies between a circular Gaussian of radius $150''$ near the center of the survey, to an elliptical Gaussian with major axis $227''$ and minor axis $150''$ towards the ends of the survey. We do not expect this to affect the polarization gradient maps we present, nor the statistical analysis that we perform in Section \ref{statgrad}.




\section{Large-scale gradient structure in the Galactic mid-plane}
\label{allsky}

\begin{figure*}
\begin{center}
\vspace{1cm}
\includegraphics[scale=0.85]{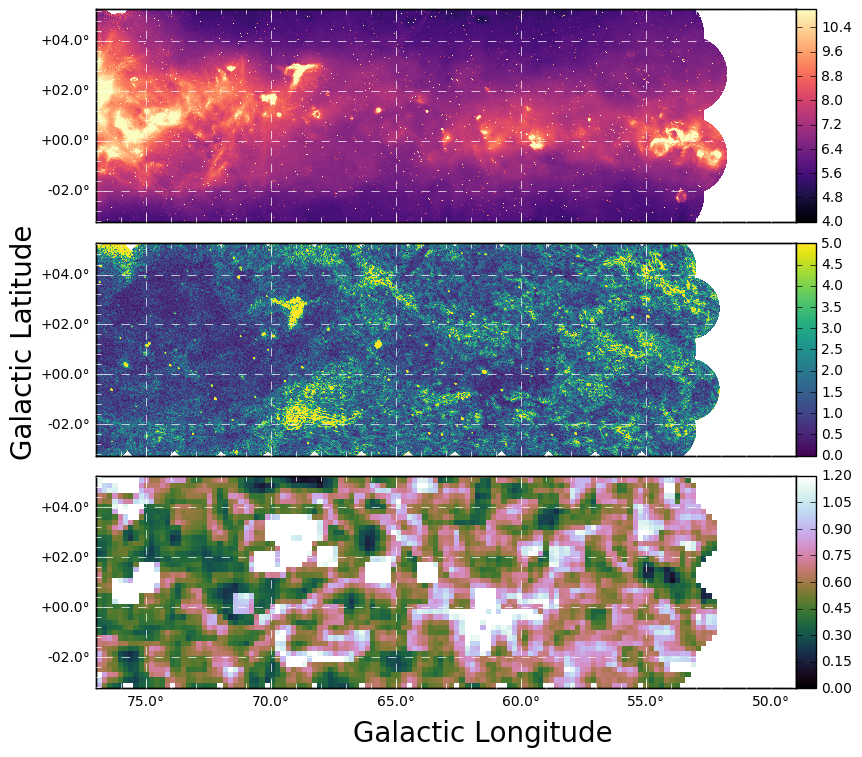}
\caption{The polarization gradient $|\nabla {\bf{P}}|$ of the CGPS at $150''$ resolution for the longitude range ${50^{\circ}}<{\ell}<{77^{\circ}}$ (blue-green, middle). 1420~MHz total-intensity (Stokes $I$) images are shown for comparison (purple-orange, top, units of K), as are the maps of the skewness of the polarization gradient (black-pink, bottom, dimensionless), calculated using an evaluation box that has $20$ beams on each side. The amplitude scale for the gradient images is the same for Figures \ref{pgrad_cgps_all1} to \ref{pgrad_cgps_all6}, expressed in K/degree, as is the amplitude scale for the skewness maps, but the scale of the $I$ images changes with longitude to accommodate the large range of brightness temperature.}
\label{pgrad_cgps_all1}
\end{center}
\end{figure*}

\begin{figure*}
\begin{center}
\vspace{1cm}
\includegraphics[scale=0.85]{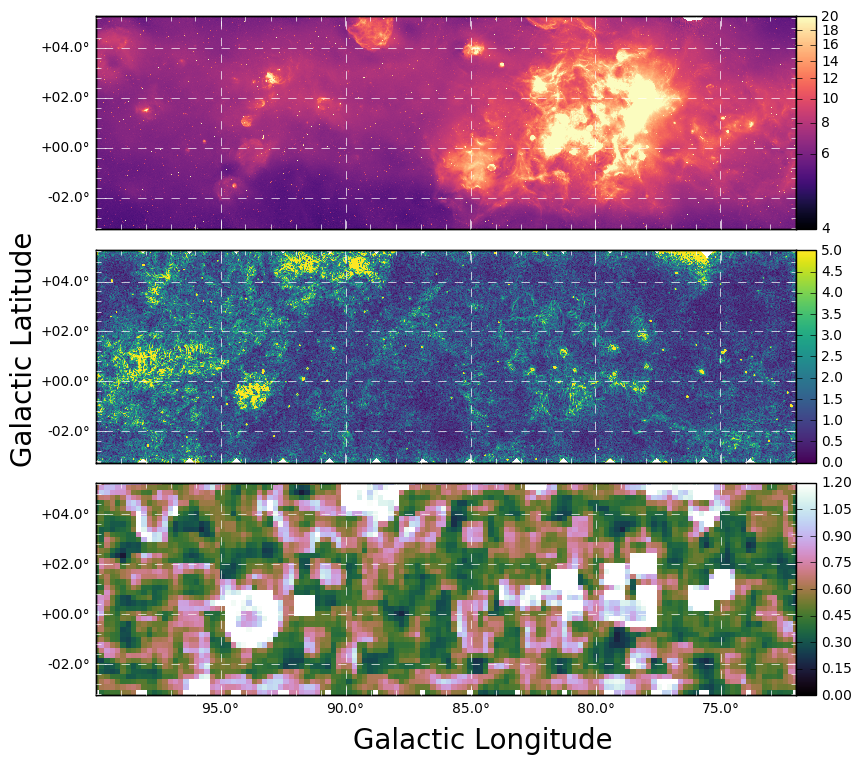}
\caption{The same as Figure \ref{pgrad_cgps_all1}, but for the longitude range ${72^{\circ}}<{\ell}<{100^{\circ}}$. A square root color scale has been applied to Stokes $I$.}
\label{pgrad_cgps_all5}
\end{center}
\end{figure*}

\begin{figure*}
\begin{center}
\vspace{1cm}
\includegraphics[scale=0.85]{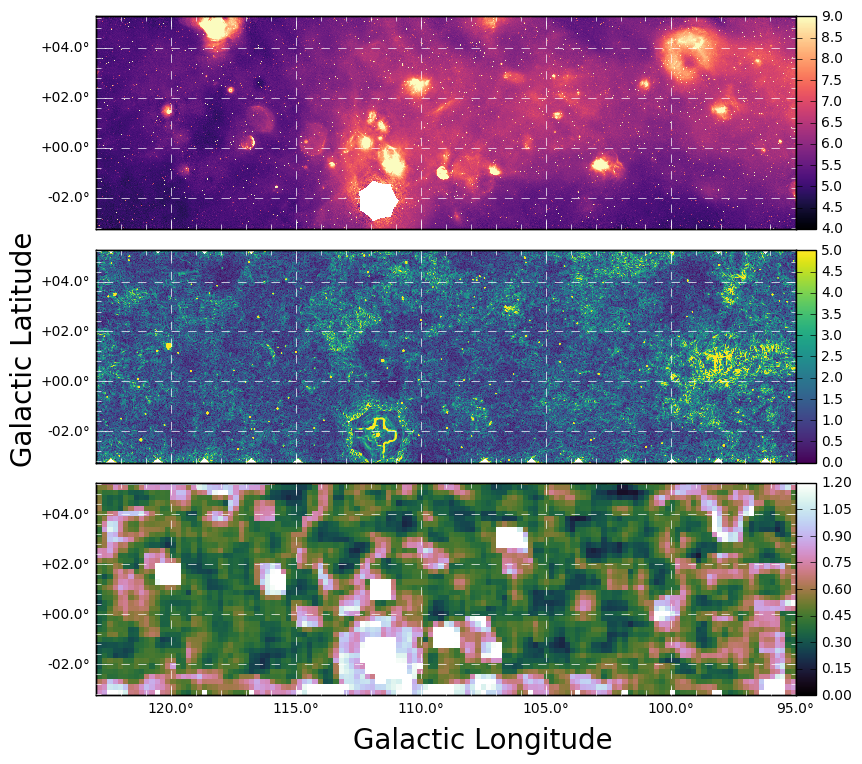}
\caption{The same as Figure \ref{pgrad_cgps_all1}, but for the longitude range ${95^{\circ}}<{\ell}<{123^{\circ}}$.}
\label{pgrad_cgps_all4}
\end{center}
\end{figure*}

\begin{figure*}
\begin{center}
\vspace{1cm}
\includegraphics[scale=0.85]{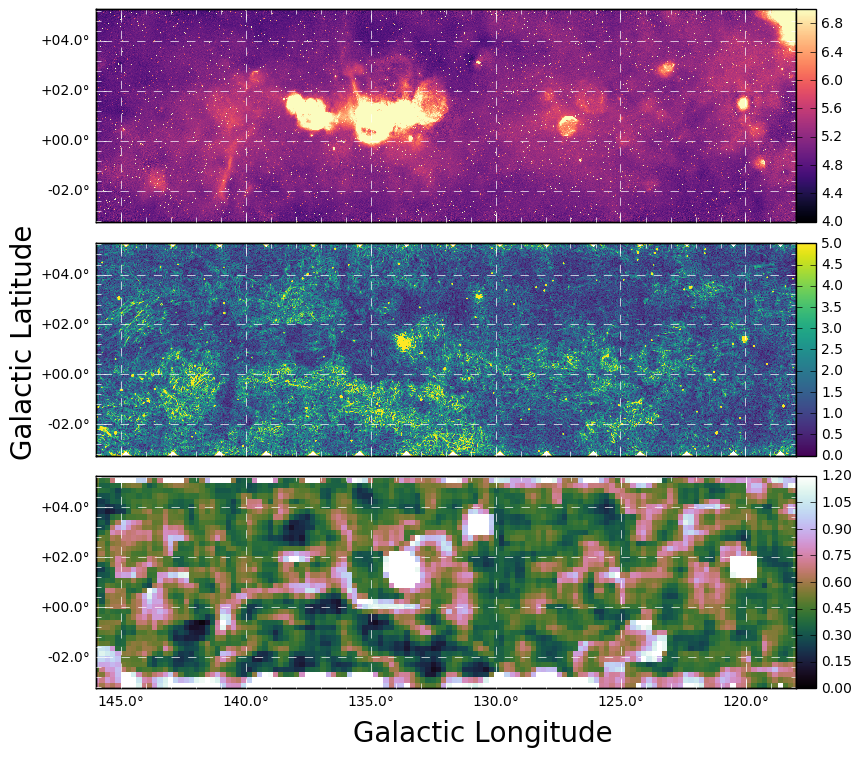}
\caption{The same as Figure \ref{pgrad_cgps_all1}, but for the longitude range ${118^{\circ}}<{\ell}<{146^{\circ}}$.}
\label{pgrad_cgps_all3}
\end{center}
\end{figure*}

\begin{figure*}
\begin{center}
\vspace{1cm}
\includegraphics[scale=0.85]{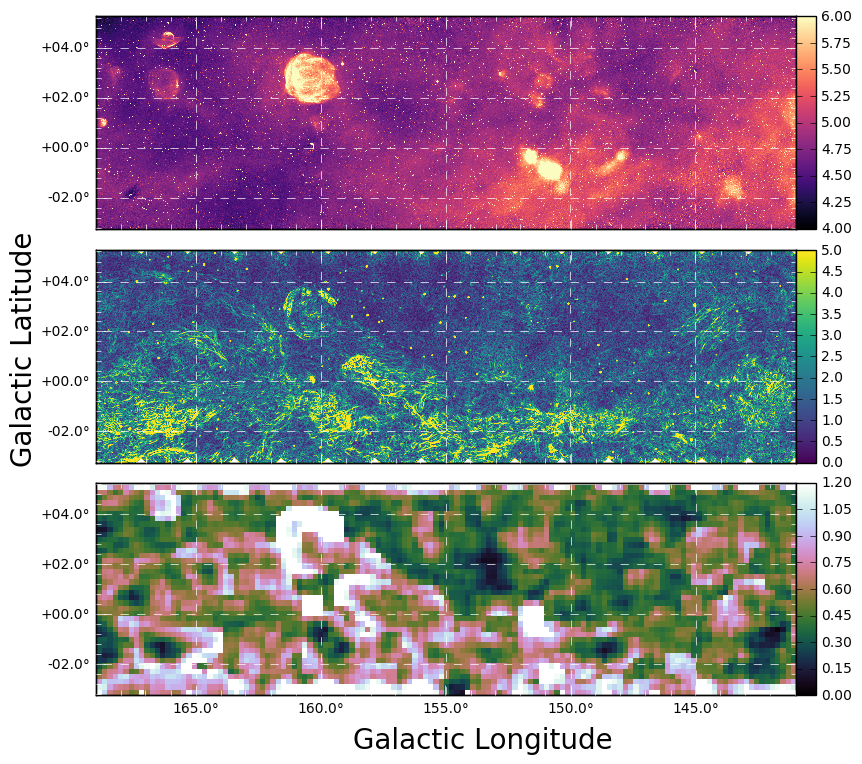}
\caption{The same as Figure \ref{pgrad_cgps_all1}, but for the longitude range ${141^{\circ}}<{\ell}<{169^{\circ}}$.}
\label{pgrad_cgps_all2}
\end{center}
\end{figure*}

\begin{figure*}
\begin{center}
\vspace{1cm}
\includegraphics[scale=0.85]{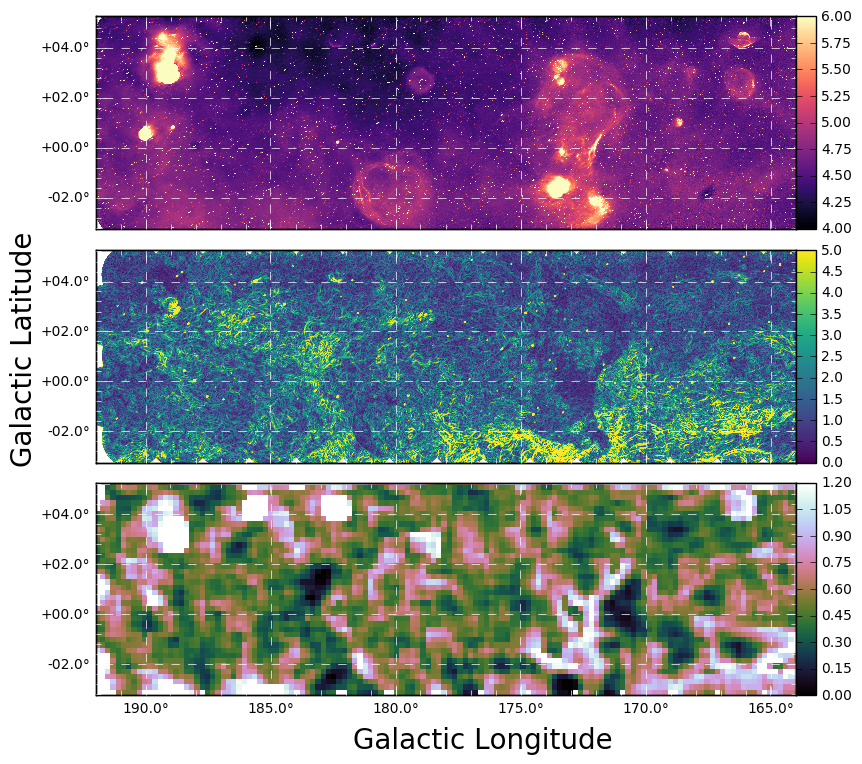}
\caption{The same as Figure \ref{pgrad_cgps_all1}, but for the longitude range ${164^{\circ}}<{\ell}<{192^{\circ}}$.}
\label{pgrad_cgps_all6}
\end{center}
\end{figure*}

Figures \ref{pgrad_cgps_all1} to \ref{pgrad_cgps_all6} show the polarization gradient image of the entire mid-plane area mapped by the CGPS, at an angular resolution of $150''$ (blue-green, middle). The CGPS total-intensity data (Stokes $I$) are shown for comparison (purple-orange, top). There is very little correlation of the $I$ image with the $|\nabla{\bf{P}}|$ image, with the exception of a few large, bright supernova remnants (SNRs) and the depolarizing effects of nearby \ion{H}{2} regions. If the polarization gradients are the products of Faraday rotation in the ISM there should be no correlation with $I$; a correlation with $I$ would indicate that the gradients are tracing the polarized emission from discrete objects, and in such locations the $|\nabla{\bf{P}}|$ image gives no information on turbulence in the ISM \citep{Iacobelli2014}.

In places there is very little small-scale polarization structure and the level of the gradient signal is also low, below the noise in the image. An example is provided by the region\footnote{This is the Fan region, which exhibits bright, smooth polarized emission, with little fine structure.}  ${145^{\circ}}\leq{\ell}\leq{159^{\circ}}, {0^{\circ}}\leq{b}\leq{5^{\circ}}$. In regions like this the hexagonal noise pattern (see Section~\ref{data}) becomes evident. We also note obvious spurious features around the position of Cas~A, (${\ell},{b})=({111.7^{\circ}},{-2.1^{\circ}})$.

We find numerous ``double-jump'' features throughout the CGPS, indicating the presence of strong shocks in the magneto-ionic material of the Galactic plane \citep{Burkhart2012}. Some ``double-jump'' features only become visible at poor angular resolution, for example at an angular resolution of $20'$ (see Figure \ref{pgradres}f, $({\ell},{b})=({112.4^{\circ}},{5.5^{\circ}})$), which may indicate relatively nearby strong shocks, whereas others are visible at a resolution of $150''$, for example at (${\ell},{b})=({112.5^{\circ}},{9.2^{\circ}})$ in Figure \ref{pgradres}. This feature can be seen in polarized intensity; see Figure 11 of \cite{Landecker2010}. \citet{Iacobelli2014} point out similar structures in the Southern sky at $2.3$ GHz. 

By analyzing the shape of the polarization gradient filaments qualitatively throughout the survey, we find that the structures in the $|\nabla{\bf{P}}|$ image tend to be longest towards the Galactic anti-center, and decrease in length towards the inner Galaxy (but we note some exceptions below). We also find that the gradient structures tend to be shorter in more depolarized regions, and hence the length of $|\nabla{\bf{P}}|$ structures may correlate with Galactic longitude due to increased depolarization along lines of sight that pass closer to the inner Galaxy. It is also possible that the small filaments are related to magneto-ionic material that is further away from us than that of long filaments, or the turbulence towards the inner Galaxy may be stronger on small scales than the turbulence towards the anti-center.


\subsection{The polarization horizon}
\label{polhor}

It is important for the interpretation of the gradient data to understand the distances to the polarization features that we see in these images, and we therefore discuss the concept of the polarization horizon, which was introduced by \citet{Landecker2002}, and further discussed by \citet{Uyaniker2003} and \citet{Kothes2004}. The polarization horizon describes a characteristic distance, $d_{\rm ph}$, beyond which polarized emission is not detectable because of the combined effects of beam and depth depolarization. Depth depolarization refers to the depolarization caused by the superposition of emission with different polarization angles along the line of sight, due to Faraday rotation of the emission as it propagates through a synchrotron emitting medium. Hence, $d_{\rm ph}$ is a function of direction, frequency of observation and angular resolution. 

\begin{figure}
\begin{center}
\includegraphics[trim={50 175 0 130} , scale=0.45,clip]{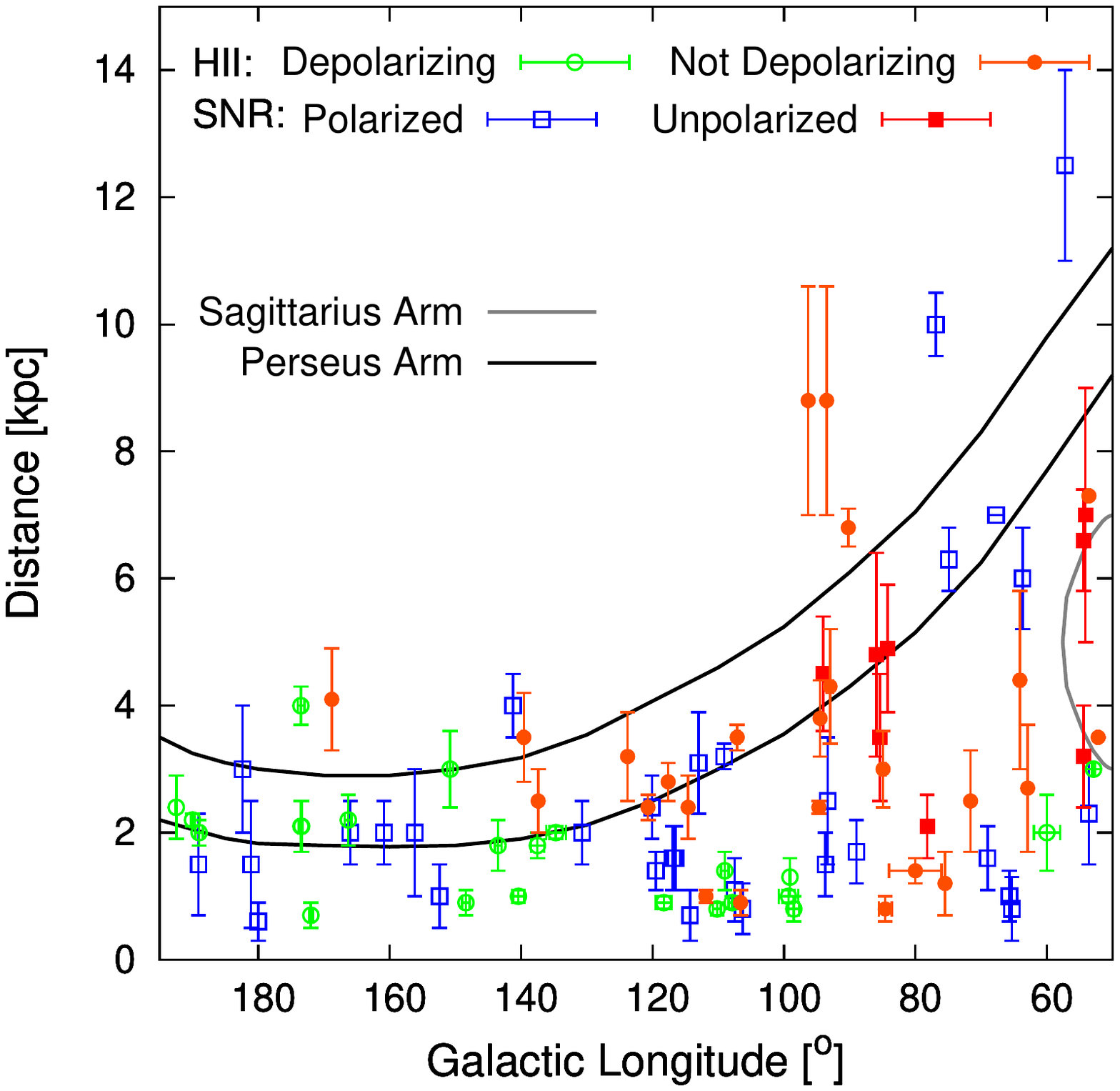}
\caption{Longitude-distance diagram of the locations of over 100 SNRs and \ion{H}{2} regions observed in the CGPS. SNRs whose polarization is detected and \ion{H}{2} regions that depolarize background emission (blue and green symbols respectively) are within the polarization horizon; the remaining objects (red symbols) are beyond it. The locations of the Perseus and Sagittarius arms are indicated. Reproduced from Kothes et al. 2016a (in prep).}
\label{PHl}
\end{center}
\end{figure}

SNRs and \ion{H}{2} regions help map out $d_{\rm ph}$ as a function of longitude. If a SNR is beyond the polarization horizon the combined effects of Faraday rotation in the intervening medium and beam depolarization will efface its polarized emission; a closer SNR will be observed as a polarized object. \ion{H}{2} regions within the polarization horizon will be seen, in contrast to their surroundings, to depolarize more distant emission; those beyond the polarization horizon will have no perceptible depolarization effect. Using CGPS data (at full resolution), Kothes et al. (2016a, in prep) have thoroughly `mapped' the polarization horizon by compiling distances to known SNRs and \ion{H}{2} regions, and results from that paper are reproduced here as Figure~\ref{PHl}. 

In the range ${53^{\circ}}<{\ell}<{140^{\circ}}$, $d_{\rm ph}$ is mostly about $3$~kpc. $d_{\rm ph}$ appears to increase to about $4$~kpc from ${\ell}={140^{\circ}}$ to ${\ell}={170^{\circ}}$. We would expect $d_{\rm ph}$ to increase towards the anti-center because of falling density of the medium and because the magnetic field is mostly orthogonal to the line of sight, reducing Faraday rotation and so reducing depth depolarization. The strong increase in observed polarized intensity in the anti-center \citep{Landecker2010} is evidence for this. However, there are not enough SNRs and \ion{H}{2} regions in the anti-center for a reliable determination of $d_{\rm ph}$ using the method described above.

There is a ``window'' through the polarization horizon between ${\ell}\approx{60^{\circ}}$ and ${\ell} \approx{80^{\circ}}$, allowing the detection of polarized emission from relatively large distances. This effect is discussed in an upcoming paper (Kothes et al. 2016a, in prep).

\subsection{The inner Galaxy off the Local arm: 
${53^{\circ}}<{\ell}<{70^{\circ}}$}
\label{inner}

Moving from ${\ell}={70^{\circ}}$ to ${\ell}={53^{\circ}}$, the density and amplitudes of gradient structures increase with decreasing longitude. The increase in $|\nabla{\bf{P}}|$ reflects an increase in polarized intensity through the same range of longitude. The polarization gradient features are smaller in this region than in other parts of the survey, possibly because the line of sight in this region moves from the edge of the Local arm at ${\ell}={70^{\circ}}$ into the interarm region (between the Local arm and the Sagittarius arm), so that the emission comes from further away. Filaments are generally a few tenths of a degree in length. Depolarization effects associated with the Local arm gradually decrease and the polarization horizon should move to larger distances (but the experimental data of Figure~\ref{PHl} do not have the resolution to show such effects). There are two interesting features in this region. One is an arc of high polarization gradient that curves from $(\ell,b)=(66^{\circ},4^{\circ})$ to $(\ell,b)=(63^{\circ},1^{\circ})$, and the other is a horizontal region of bright polarization gradient at $(\ell,b)=(69^{\circ},-2^{\circ},)$. Within the latter region there is a remarkable patch of gradient structure, showing many parallel filaments, some as long as 1.5 degrees. Both of these features have counterparts in polarized intensity, but not in total intensity. An investigation of the origin of these features is beyond the scope of this paper. 

\subsection{The Local arm and Cygnus X: ${70^{\circ}}<{\ell}<{85^{\circ}}$}
\label{local}

In the range ${70^{\circ}}<{\ell}<{85^{\circ}}$, lines of sight pass for distances of a few kpc along the Local arm.  Over this range of longitudes, the polarization gradient map shows only low-level features. Significant small-scale turbulence driven by star-formation activity is expected within a spiral arm, and may even be below the resolution limit, leading to strong depolarization and bringing the polarization horizon very close.

Cygnus X lies within this area (${76^{\circ}}<{\ell}<{83^{\circ}}, {-2^{\circ}}<{b}<{3^{\circ}}$). Cygnus X is a very powerful total-intensity source, and intense \ion{H}{2} regions and SNR emitters of small diameter are buried within it. The stronger of these small-diameter sources generate spurious polarized signal, and the correction for instrumental polarization does not completely remove their effects. Some of these regions are identified in Table \ref{tab:source_list}. Cygnus X is one of the closest regions of massive star formation at distances between 1 and 2.5~kpc (\citealp{Gottschalk2012}, \citealp{Rygl2012}). It includes the Cygnus OB2 association comprising $\sim{2600}$ OB stars \citep{Knodlseder2000}.  We adopt the distance of $1.40{\pm}0.08$~kpc, determined by \citet{Rygl2012} from parallax measurements of maser sources, as the distance to the whole complex. The very large amount of ionized gas within Cygnus X effectively erases all polarized signal from behind it, placing the polarization horizon on the near side of the complex. 

Low-level polarized signal is present in the direction of Cygnus X, and this must arise in the space between Cygnus X and the observer. If the few gradient structures seen at our chosen resolution ($150''$) are at a distance of 1.4~kpc, then the characteristic size of the turbulence features in this direction is about 1~pc.

\subsection{The interarm region between the Local and Perseus arms:
${85^{\circ}}<{\ell}<{105^{\circ}}$}
\label{localpers}

In the longitude range from $85^{\circ}$ to $105^{\circ}$, lines of sight pass into the inter-arm space between the Local and Perseus arms. The polarization horizon barely reaches into the Perseus arm here, so we are seeing primarily the ISM in the interarm. Gradient filaments are bright and some reach lengths of 0.5 degrees. The appearance of $|\nabla{\bf{P}}|$ closely parallels that between ${70^{\circ}}$ and ${53^{\circ}}$ and we believe that the same polarization effects apply as described in Section \ref{inner}.

\subsection{The Perseus arm: ${105^{\circ}}<{\ell}<{140^{\circ}}$}
\label{pers}

In the range ${105^{\circ}}<{\ell}<{125^{\circ}}$, lines of sight encounter star-forming regions in the Perseus arm and the distance through the inter-arm region is continuously increasing with increasing longitude. Gradient filaments are a few tenths of a degree in length here. In the longitude range ${125^{\circ}}<{\ell}<{140^{\circ}}$ there are fewer \ion{H}{2} regions than for ${105^{\circ}}<{\ell}<{125^{\circ}}$, and lines of sight penetrate further into the Perseus arm. The exception is the large \ion{H}{2} complex W3/W4/W5 at longitudes $133^{\circ}$ to $138^{\circ}$. It lies at a distance of 2.0~kpc \citep{Xu2006} on the near side of the Perseus arm, and effectively defines the polarization horizon. All polarized emission detected in the direction of W3/W4/W5 must arise in the Local arm or the intervening inter-arm, and this emission shows little or no gradient structure. The exception is a bright spot of spurious signal generated by W3, an intense compact \ion{H}{2} region (see Table \ref{tab:source_list}).

\subsection{The anti-center region: ${\ell}>{140^{\circ}}$}
\label{antic}

Beyond ${\ell}\approx{140^{\circ}}$ lines of sight are directed at the outer Galaxy and the anti-center. Polarized intensity and $|\nabla{\bf{P}}|$ increase across this region. This increase is seen at first in the southern half of the field. Between longitudes $140^{\circ}$ and $160^{\circ}$ we find strongly polarized but very smooth emission from the Fan region (\citealp{Landecker2010}; this emission extends to higher latitudes, beyond the range of the CGPS). There is no discernible gradient structure in the Fan region. The brightest gradient structures that we have detected are in the anti-center; some are as long as $1$ degree. These structures appear to have a different morphology to the polarization gradient structures observed towards the inner Galaxy. In particular, the polarization gradient shows resolved filaments towards the anti-center (e.g. near $\ell = 165^{\circ}$), but smaller-scale, less elongated features towards the inner Galaxy (e.g. near $\ell = 60^{\circ}$). We also note features in the polarization gradient that follow parts of circular arcs for ${161^{\circ}}<{\ell}<{170^{\circ}}$, which are likely related to a stellar wind bubble (to be discussed by Kothes et al. 2016b, in prep).

\subsection{The Northern Latitude Extension}
\label{nle}

\begin{figure*}
\begin{center}
\includegraphics[scale=0.79]{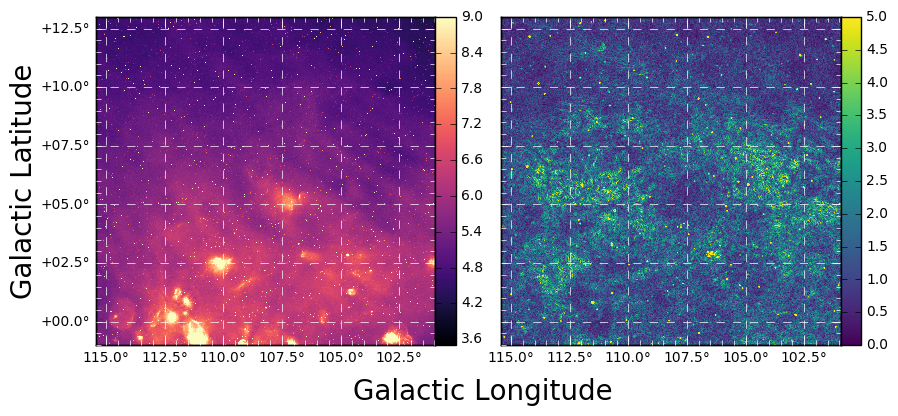}
\caption{The total intensity map at the full resolution of the CGPS in K (left), and the corresponding polarization gradient of the northern latitude extension of the CGPS at $150''$ resolution in K/degree (right).}
\label{pgrad_NLE}
\end{center}
\end{figure*}

The Northern Latitude Extension of the CGPS covers the area $101^\circ<l<116^\circ$ and $5^\circ<b<17.5^\circ$. We present the polarization gradients in this area for $b<12^\circ$ in Figure~\ref{pgrad_NLE} (there is little polarization gradient structure for $b>12^\circ$ at this resolution). As discussed by \citet{Landecker2010}, there is a transition in the appearance of polarized intensity between ${b}={8^{\circ}}$ and ${b}={10.5^{\circ}}$: below this region the sizes of polarization structures are $\sim{3'}$ and above it sizes are $\sim{20'}$.  This is identified by \citet{Landecker2010} as the transition from the Galactic disk to the halo at the top of the Perseus arm; this interface is also seen in $|\nabla{\bf{P}}|$. Below ${b}\approx{8^{\circ}}$, $|\nabla{\bf{P}}|$ has much the same appearance as it has at mid-plane in the outer Galaxy. Above ${b}\approx{9^{\circ}}$ significant gradient signal is found only in patches and the hexagonal pattern produced by non-uniform image noise becomes obvious. However, gradient maps produced with an angular resolution above $210''$ show gradient structures up to $b \approx 14^{\circ}$, which may either be nearby features in the Local Arm, or large-scale features in the Galactic halo.  A small gradient feature at $({{\ell},b})=({111^{\circ}, 11.6^{\circ}})$ is attributable to the planetary nebula DeHT~5, discussed by \citet{Ransom2010}.

\citet{Landecker2010} identified three finger-like polarization features, seen in polarized intensity and in polarization angle near $({\ell,b})=({112.5^{\circ},{9.2^{\circ}}})$ (Figure 11 in that paper). They are clearly detected as gradient features with a ``double-jump'' structure, believed to indicate a strong shock.

\subsection{Gradient signatures of individual objects}
\label{objects}

Identifiable objects, such as SNRs and \ion{H}{2} regions, create recognizable and quite strong effects in the gradient images and many are quite obvious in Figures~\ref{pgrad_cgps_all1} to \ref{pgrad_cgps_all6}. Examination of these effects will be the subject of a future paper. Some prominent features that are evident in Stokes I are listed in Table \ref{tab:source_list}. 

\begin{table}
\centering
\caption{Locations of prominent features in the Stokes I images that may influence polarization gradient images.} \label{tab:source_list}
\begin{tabular}{| c c | c |}
\hline
\hline
$\ell$  & $b$ & Comments\\
(degrees) & (degrees) & \\
\hline
$180.0$ & $-1.7$  & SNR S147, diameter $\approx 3^{\circ}$ \\
$173.0$ & $-0.3$  & \ion{H}{2} region complex\\
& & Sharpless 229/232/234/235/236 \\
$166.3$ & $4.0$ & SNR VRO 42.05.01\\
$160.9$ & $2.6$  & SNR HB9 \\
$133.8$ & $1.2$ & Compact \ion{H}{2} region W3\\
& & spurious instrumental effect\\
$130.7$ & $3.1$  & 3C58 - polarized SNR\\
& & produces polarization artefact\\
$120.1$ & $1.4$ & 3C10 (Tycho's SNR)\\
& & produces polarization artefact\\
$111.7$ & $-2.1$ & Cas A, strong sidelobe effects\\
$93.8$ & $-0.5$ & SNR CTB104A\\
$89.0$ & $4.1$ & SNR HB21\\
$81.3$ & $1.0$ & \ion{H}{2} region DR17\\
& & spurious instrumental effect\\
$79.3$ & $1.3$ & \ion{H}{2} region DR7\\
& & spurious instrumental effect\\
$79.3$ & $0.3$  & \ion{H}{2} region DR15\\
& & spurious instrumental effect\\
$78.2$ & $1.8$ & Bright part of SNR G78.2+2.1\\
& & spurious instrumental effect\\
$78.0$ & $0.6$ & \ion{H}{2} region DR6\\
& & spurious instrumental effect\\
$76.0$ & $4.5$ & Sidelobe effects from Cyg A \\
$75.8$ & $0.4$ & \ion{H}{2} region ON2\\
& & spurious instrumental effect  \\
$69.0$ & $2.7$ & SNR CTB80\\
\hline
\end{tabular}
\end{table}

\section{Statistical Analysis of Polarization Gradients}
\label{statgrad}
\subsection{Method to Mask Point Sources} \label{mask}
To calculate the skewness of the polarization gradient maps, we used a sliding-box method, calculating the skewness of the polarization gradient pixels that fall within the evaluation box. Many polarized point sources are visible in the CGPS, and these appear as bright sources in the polarization gradient images. They significantly distort the PDF of the polarization gradient and adversely influence calculations of the skewness. To remove the influence of the polarized point sources, we used the following masking procedure:

\begin{enumerate}
\item We ran the Background And Noise Estimation (BANE) tool and Aegean source finding program \citep{Hancock2012} on the Stokes $Q$ and $U$ mosaics. The source finding was conducted so that both positive and negative point sources in the $Q$ and $U$ maps were detected. The source lists obtained for Stokes $Q$ and $U$ were then combined, to create a list of all polarized sources.


\item From the source list obtained, the sources in the $Q$ and $U$ maps were masked. Two masks were created, one for which sources were masked to a radial extent of $2.5$ standard deviations of the fitted Gaussian, and another where sources are masked to a radial extent of $5$ standard deviations of the fitted Gaussian. These masks were created for the mosaic of the mid-plane of the CGPS, and for the mosaic of the northern latitude extension.

\item The mask created using a radial extent of $2.5$ standard deviations was applied to the $Q$ and $U$ mosaics, ensuring that no polarized point sources appear in either mosaic. We then smoothed the masked mosaics of Stokes $Q$ and $U$ to the same resolutions as in Section \ref{resstudy}, using the \texttt{convol} task of \texttt{MIRIAD} \citep{Sault1995}. 

The \texttt{convol} task does not take masked pixels into consideration, and this causes the intensity of the pixels near a masked source to decrease towards zero, with the effect being stronger closer to the masked source. This creates a gradient in $Q$ and $U$ around each masked source, which would appear as a bright ring around the mask in mosaics of the polarization gradient.

\item To ensure that polarization gradient rings are not present in the final gradient maps produced, we applied the second mask, created using a radial extent of $5$ standard deviations, to the smoothed mosaics of Stokes $Q$ and $U$. 

\item Finally, we produced mosaics of the polarization intensity and the polarization gradient for each angular resolution. 

\end{enumerate} 

\begin{figure*}
\begin{center}
\includegraphics[scale=0.79]{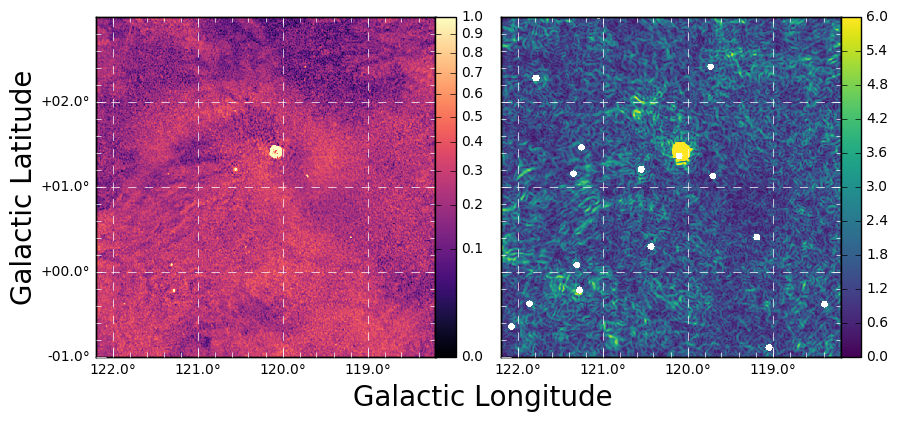}
\caption{The polarized intensity (left) in K, and the masked polarization gradient (right) at $150''$ resolution in K/degree, for a section of the CGPS. Some masked sources are weak, and difficult to see in polarized intensity. Tycho's supernova remnant is visible towards the center of the image.}
\label{masked_fig}
\end{center}
\end{figure*}

We show an example of the masked mosaic of the polarization gradient in Figure \ref{masked_fig}, at an angular resolution of $150''$, compared to the unmasked polarized intensity image of the same area (units in K). The image of the polarization gradient demonstrates that applying a second mask has removed the polarization gradient rings from the mosaic, and masked all polarized point sources. 

We have considered the possibility of only masking out point sources after smoothing has been performed, however we believe that masking point sources before the smoothing procedure ensures that polarized radiation from the sources will not influence the polarization gradient maps. This is particularly important for the polarization gradient maps produced for large smoothing scales, as otherwise the sources may protrude from the applied masks.

Although polarized point sources and polarization gradient rings have been removed from the polarization gradient maps, discrete, extended sources of polarized emission, such as SNRs, are still present (see Table \ref{tab:source_list} and Figure \ref{masked_fig}). These discrete objects will affect the calculation of the skewness of the polarization gradient, which we discuss in the following section.

\subsection{Calculation of Skewness Maps} \label{skewmaps}
To produce images of the skewness of the polarization gradient, we use a sliding box method, where the skewness of the polarization gradient is calculated using all pixels that fall within a specified box. This box is moved over a rectangular grid of evaluation points, to sample all of the pixels in the mosaics. Two competing factors constrain the ideal size of the box to use to calculate the skewness. If a large box is used to calculate the skewness, then this would ensure that the underlying PDF of the polarization gradient is well sampled, and the skewness accurately measured. However, this degrades the resolution of the image of the skewness of the polarization gradient, and so a small box may be preferable to retain as much spatial information as possible. 

To examine the effect that the size of the box has on the skewness map produced, we calculated skewness maps using boxes whose sides were $20$, $40$, $80$ and $120$ times the angular resolution of the mosaic. The number of independent data points used to calculate the skewness in each of these cases is $400$, $1600$, $6400$ and $14400$ respectively, based on the number of telescope beams that fit within the box.

To construct a grid of evaluation points, we space evaluation points by one quarter of the box size, so that we sample the polarization gradient images at a rate above the Nyquist rate. Mosaics smoothed to poorer angular resolution have fewer evaluations of the skewness, because the grid spacing depends on the angular resolution.

\begin{figure*}
\begin{center}
\includegraphics[scale=0.79]{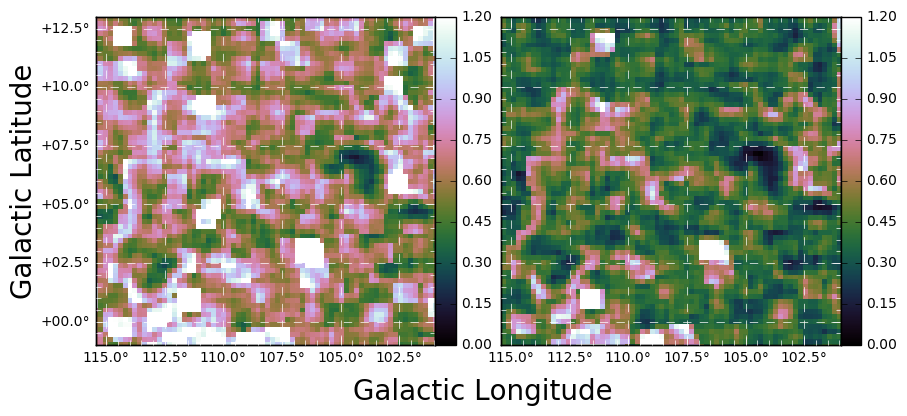}
\caption{The skewness (dimensionless) of the polarization gradient map of the northern latitude extension that has been smoothed to $150''$ resolution, using a box with $20$ beams on each side. The skewness map on the left was produced without truncation, and the skewness map on the right was produced after removing the top $1\%$ of polarization gradient values from the PDF.}
\label{skew_trunc}
\end{center}
\end{figure*}

Finally, to reduce the influence of polarization gradient rings and polarized discrete objects on the calculated skewness maps, we truncate the PDF of polarization gradient values within the evaluation box, so that the top $1\%$ of polarization gradient values are removed from the PDF before the skewness is calculated. As this step modifies the PDF, we consider whether it will substantially change the measured skewness values, and the features observed. In Figure \ref{skew_trunc} we compare the effect of this truncation on the skewness of the polarization gradient maps of the northern latitude extension, smoothed to a final resolution of $150''$. The skewness image on the left was obtained without truncation, and the image on the right was obtained with truncation.

We find that performing the truncation of the PDF removes some bright, square-shaped regions from the skewness map, indicating that the influence of polarized discrete objects has been reduced. We also find that the features of the skewness map are not affected by the truncation, and that performing the truncation tends to reduce the skewness values by approximately $0.2$. Hence, the truncation should not affect any conclusions that we might draw from these skewness maps, as the features in the skewness maps remain intact, and a change in skewness of $0.2$ would not affect the regime of turbulence that is implied by the value of the skewness (see Figure 7 of \citealt{Burkhart2012}).

\subsection{Qualitative Analysis of Skewness of $|\nabla{\bf{P}}|$} \label{SkewAnalysis}

In Figures \ref{pgrad_cgps_all1} to \ref{pgrad_cgps_all6} we compare the $150''$ polarization gradient map (blue-green, middle), for the mid-plane of the CGPS, to the skewness of this map (green-pink, bottom) produced using a box that has $20$ beams on each side. The dominant conclusion from the skewness images in Figures \ref{pgrad_cgps_all1} to \ref{pgrad_cgps_all6} is that the skewness of the polarization gradient appears to be largest around the edges of bright polarization gradient regions, rather than inside regions of bright polarization gradients, or regions that have a `double-jump' feature (that indicate a strong shock). 

For example, near ${\ell}={159^{\circ}}$, the skewness is largest on the boundaries of the bright polarization gradient regions in this direction, and small within the bright polarization gradient regions. As bright polarization gradients signify large changes in the electron density or the magnetic field along a line of sight, they should be related to high sonic Mach number turbulence, and high values of the skewness, but the opposite is observed. A more extreme example can be found for ${172^{\circ}}<{\ell}<{174^{\circ}}$, where there is a bright \ion{H}{2} region that depolarizes all emission from behind it, and around this region there are bright polarization gradient features.  As the foreground \ion{H}{2} region is unrelated to the turbulence in the ISM, the skewness of the polarization gradient is not probing the turbulence in this direction. 

To confirm that the skewness is properly measuring the skew of the underlying PDF of the polarization gradient, we examined the PDFs of the polarization gradient over a horizontal strip that cuts across the high skewness values at ${172^{\circ}}<{\ell}<{174^{\circ}}$. As the calculation box moves from the \ion{H}{2} region to the bright polarization gradient region, the PDF becomes slightly more skewed because high polarization gradient values are added to the tail of the distribution. As the box moves further into the bright polarization gradient region, this tail becomes less pronounced because the average polarization gradient in the box has risen, and this causes a decrease in skewness. We believe that the skewness is tracing subtle changes in the PDF of the polarization gradient.



\begin{figure*}
\begin{center}
\includegraphics[scale=0.8]{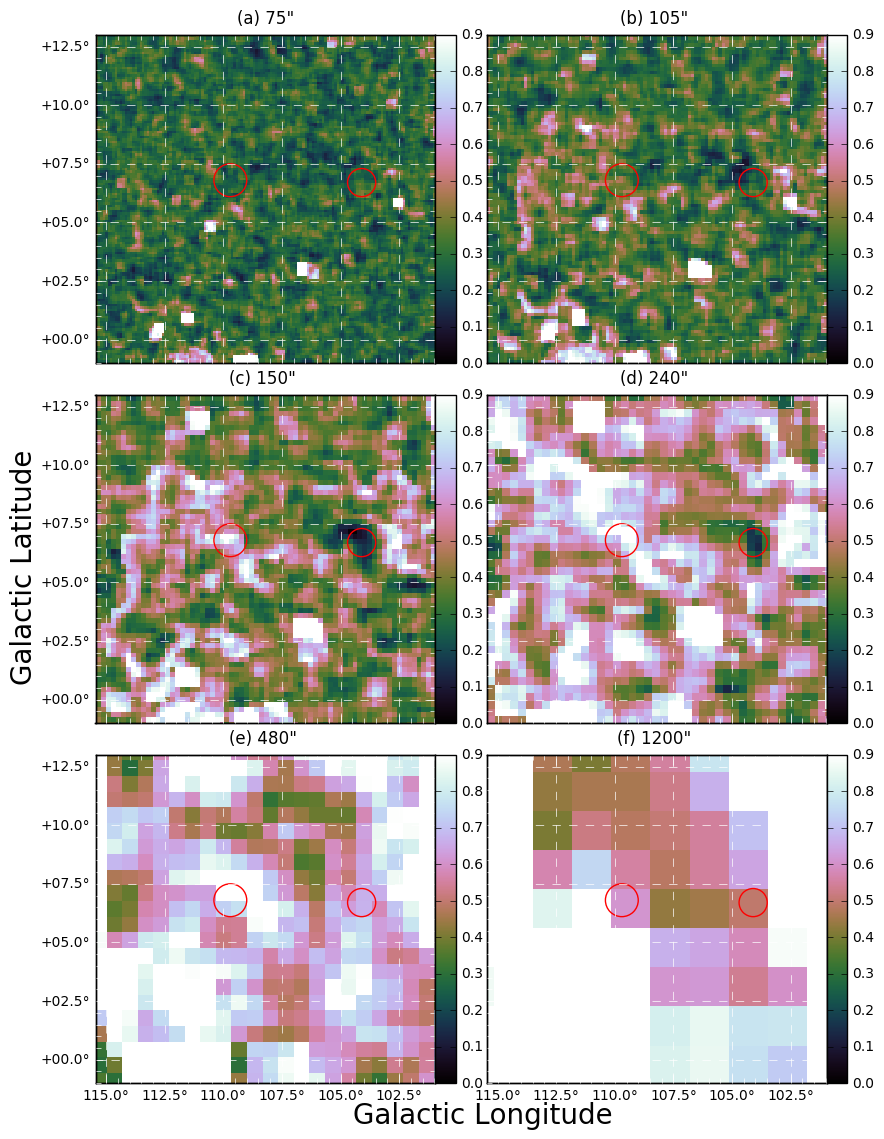}
\caption{The skewness (dimensionless) of the polarization gradient for the northern latitude extension of the CGPS, for various angular resolutions of the polarization gradient map, calculated using $20$ beams on each side of the evaluation box. The angular resolutions are a) 75'', b) 105'', c) 150'', d) 240'', e) 480'', f) 1200''. The colorbars give the magnitude of the skewness (unitless). The red circles denote the areas used to study how the skewness depends on angular resolution.}
\label{skewres}
\end{center}
\end{figure*}

We investigated whether the skewness acts as an edge detection algorithm at other resolutions by comparing skewness maps produced using polarization gradient maps at different angular resolutions, in each case using $20$ beams on each side of the evaluation box. In Figure \ref{skewres} we show the skewness of the polarization gradient for the northern latitude extension of the CGPS, for the same angular resolutions as Figure \ref{pgradres}. 

We find that there is an apparent increase in the skewness of the polarization gradient as the angular resolution becomes poorer. To understand the cause of this, we studied how the skewness of the polarization gradient changes with increasing smoothing scale in two small circular areas (shown as red circles in Figure \ref{skewres}). One area was centered at $(\ell,b) = (104.1^\circ,6.7^\circ)$ with a radius of $0.6^{\circ}$, and another was placed at $(\ell,b) = (109.7^\circ,6.8^\circ)$ with a radius of $0.7^{\circ}$. The former was chosen because it has low skewness, and is within a bright polarization gradient region, and the latter was chosen because it has high skewness, and is located on the edge of a bright polarization gradient region. At the position of low skewness, we found that the skewness initially decreased as the smoothing scale increased, but then increased monotonically. At the position of high skewness, the skewness initially increased with increasing smoothing scale, and then decreased. 

These observations can be explained in terms of how the size of the evaluation box increases with increasing smoothing scale. If nearby edges of the bright polarization gradient regions become enclosed by the evaluation box as the box grows, then the skewness will increase. Otherwise, the skewness will decrease, as increasing the smoothing scale decreases the variations in Stokes $Q$ and $U$, and hence decreases the largest values of the polarization gradient, so that the tail of the $|\nabla{\bf{P}}|$ PDF is less pronounced. 

We find that at every angular resolution we study, the skewness of the polarization gradient is influenced by edge effects, and edge effects may be more prominent for poor angular resolution. 

\begin{figure*}
\begin{center}
\vspace{1cm}
\includegraphics[scale=0.85]{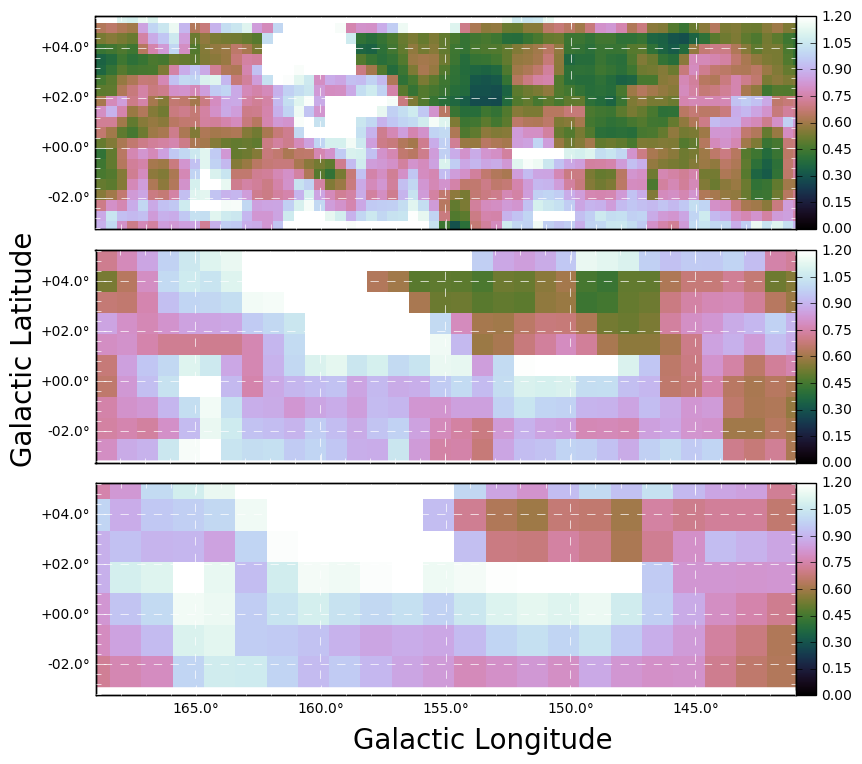}
\caption{The skewness (dimensionless) of the polarization gradient $|\nabla {\bf{P}}|$ over the longitude range ${141^{\circ}}<{\ell}<{169^{\circ}}$ at $150''$ resolution. The skewness maps were produced using evaluation boxes with $40$ (top), $80$ (middle) and $120$ (bottom) beams across the width of the box.}
\label{skew_box_comp}
\end{center}
\end{figure*}

It is possible that an evaluation box with $20$ beams on each side is too small to properly probe the turbulence, and so we compared the skewness maps produced using boxes with widths of $40$, $80$ and $120$ beams, as shown in Figure \ref{skew_box_comp}, for ${141^{\circ}}<{\ell}<{169^{\circ}}$. We find that as the size of the evaluation box increases, the mean skewness in this portion of the Galactic plane increases, and the structures in the skewness map produced using a box with a width of $20$ beams simply become blurred. Similar to the results found for different angular resolutions, this implies that as the box size increases, more edges of polarization gradient structures are included within the box, causing the value of the skewness to increase. This implies that increasing the size of the evaluation box does not provide a more accurate measurement of the skewness, as the skewness is too sensitive to the number and magnitude of the inhomogeneities within the box.

To confirm that these problems with the skewness are not a result of how the CGPS data have been reduced and analyzed, we applied our sliding box method to the polarization gradient image of the Southern Galactic Plane Survey test region, produced by \cite{Gaensler2011}. For this image, we again find that the skewness of the polarization gradient is largest around the edges of bright polarization gradient regions, and that the smallest skewness values tend to be found in regions where there are bright polarization gradient filaments.

We conclude that an intrinsic property of the skewness as a statistic is that it behaves as an edge detecting algorithm when applied to images. As the edges of bright polarization gradient regions are unrelated to the turbulence that is revealed by the polarization gradient method, this implies that the skewness of the polarization gradient cannot probe the turbulence observed in large portions of the sky. 

This is true not only for regions of high skewness located near well-defined edges, but also for regions of medium skewness, where edges that have a small contrast between high and low polarization gradient may provide an unknown contribution to the measured skewness value. In these regions, there will always be uncertainty as to how much the interstellar turbulence revealed by the polarization gradients contributes to the measured skewness, making the skewness an unreliable probe of interstellar turbulence.

There are several reasons why the correlation between the skewness of the polarization gradient and the sonic Mach number, found for the simulations by \cite{Burkhart2012}, may not be applicable here. One reason is that the magnitude of the polarization gradient is lower in areas coincident with foreground depolarizing gas. This leads to a higher skewness on the boundary of the depolarizing gas, which is not indicative of the regime of turbulence. An extreme example is the \ion{H}{2} region at ${172^{\circ}}<{\ell}<{174^{\circ}}$, but there may be more diffuse depolarizing clouds throughout the Galactic plane, that have a more subtle effect on the skewness. Another reason is that \cite{Burkhart2012} considered the scenario where the turbulent medium is illuminated from behind by polarized emission. However, we do not currently know whether the correlation between skewness and sonic Mach number in their simulations will persist for the scenario where polarized emission comes from within the turbulent medium, and this scenario may be prevalent in the Galactic plane. Finally, \cite{Burkhart2012} simulate homogeneous turbulence, where the driving mechanism of the turbulence is the same throughout the simulation cube. In the interstellar medium, inhomogeneous turbulence is likely to dominate, and the skewness of the polarization gradient may be large along the edges of homogeneous regions.

\subsection{Quantitative Analysis of Skewness and Kurtosis of $|\nabla{\bf{P}}|$} \label{QuanSkew}
\begin{figure*}
\begin{center}
\includegraphics[trim={10 0 0 0} , scale=0.8,clip]{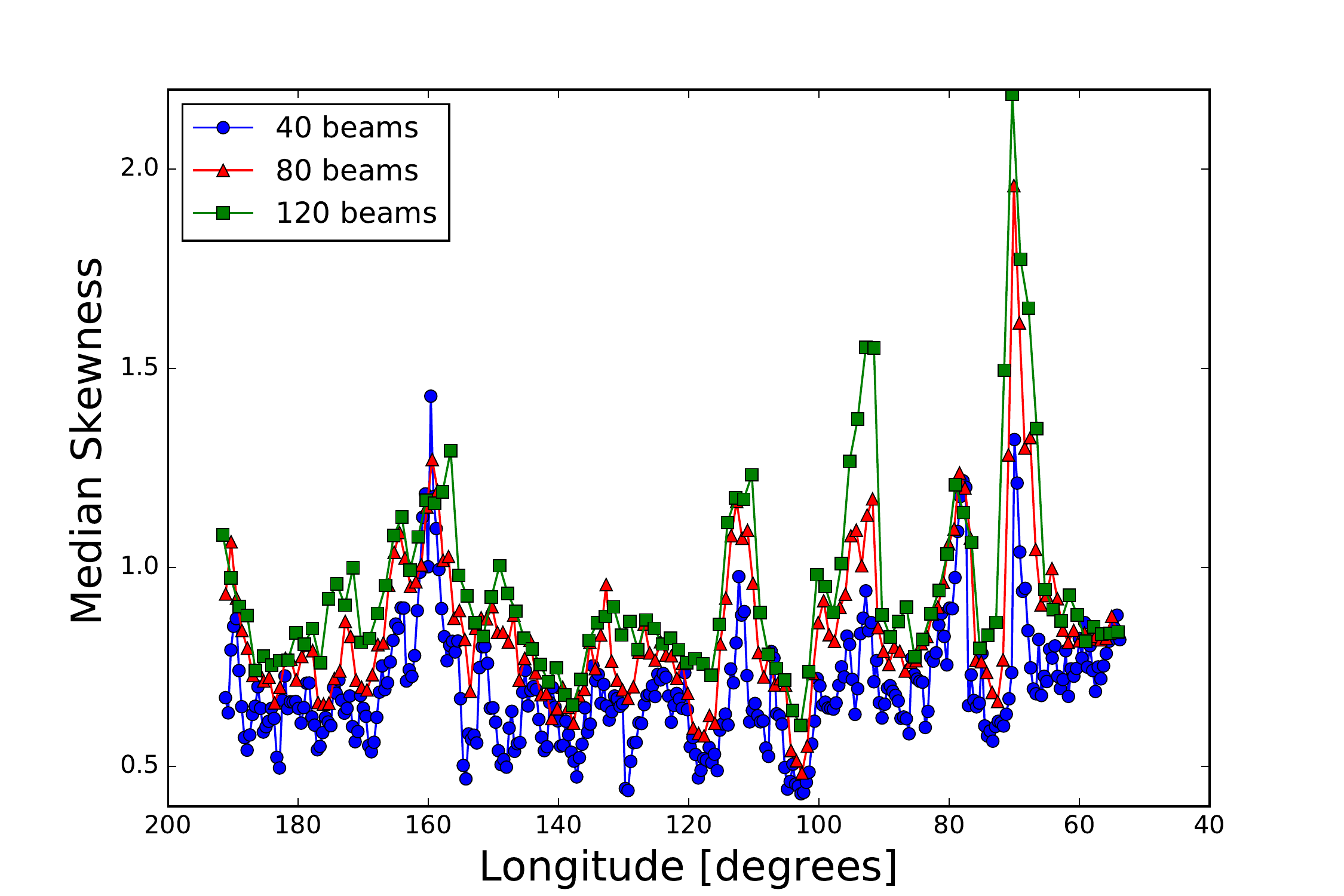}
\caption{The median skewness (dimensionless) of the polarization gradient map at $150''$ resolution, measured at each longitude, for different sizes of the evaluation box used to calculate the skewness. The median skewness measured using an evaluation box with a width of $40$, $80$ and $120$ beams are shown as blue circles, red triangles, and green squares respectively.}
\label{skew_vs_long}
\end{center}
\end{figure*}

To study whether the number or magnitude of the inhomogeneities probed by the skewness varies throughout the survey, we plot the skewness of the polarization gradient map for the Galactic plane mosaic, produced at $150''$ resolution, as a function of Galactic longitude, as shown in Figure \ref{skew_vs_long}. Each data point represents the median skewness at that longitude, calculated for evaluation boxes with $40$ (blue circles), $80$ (red triangles) and $120$ (green squares) beams across the width of the box.

There are some peaks that can be seen in this plot, which are present for all of the box sizes that we study. These peaks correspond to strong polarized continuum emitters such as supernova remnants, or are instrumental polarization artefacts generated by very bright HII regions; the peaks do not denote areas of strong turbulence in the ISM. Away from these peaks, we find that the median skewness is approximately independent of Galactic longitude for each of the box sizes that we study. In particular, there are no significant differences in the skewness values obtained for the regions of the Galactic plane that are discussed in Section \ref{allsky}. We also find that the median skewness tends to increase as the width of the box increases, for the entire portion of the Galactic plane covered by the CGPS data.

As the skewness of the polarization gradient appears to be more sensitive to the size of the evaluation box than to the structures in the polarization gradient map, this strengthens our belief that the skewness is primarily probing the number and magnitude of the inhomogeneities within the evaluation box, rather than probing the underlying turbulence. Hence, we find that the skewness of the polarization gradient is not a useful statistic for probing observed magnetized interstellar turbulence.

\begin{figure*}
\begin{center}
\includegraphics[scale=0.8]{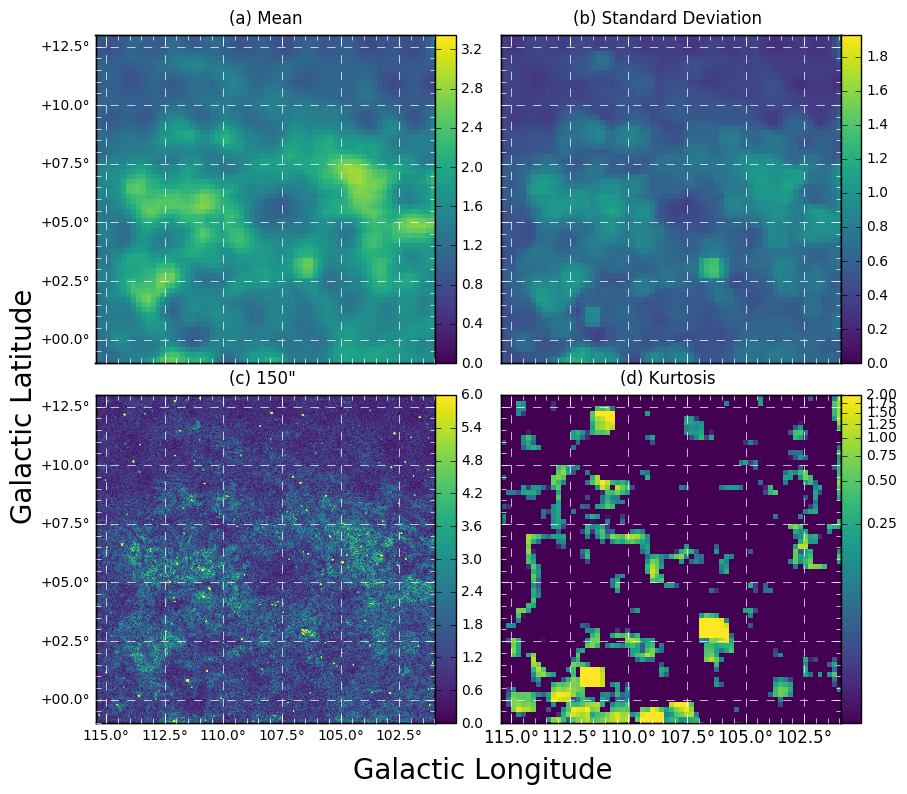}
\caption{The a) mean (K/degree), b) standard deviation (K/degree), and d) kurtosis (dimensionless) of the polarization gradient map of the northern latitude extension of the CGPS, c), smoothed to $150''$ resolution. A logarithmic color scale is used for the kurtosis.}
\label{other_stats}
\end{center}
\end{figure*}

To see whether other moments of the PDF are similarly affected by inhomogeneities, we use our sliding box method to calculate the mean, standard deviation, and kurtosis of the polarization gradient. These moments are shown in Figure \ref{other_stats}, for the polarization gradient image of the northern latitude extension, at an angular resolution of $150''$, using an evaluation box with $20$ beams on each side. We find that the mean and standard deviation do appear to trace the polarization gradient structures, and hence may be better suited to probing turbulence than the skewness. In particular, Figure 7 of \cite{Burkhart2012} shows that both the mean and standard deviation of the polarization gradient depend on the sonic Mach number, and so these statistics may prove useful. This is expected, as stronger turbulence should produce larger variations in Stokes $Q$ and $U$, and hence larger values of $|\nabla{\bf{P}}|$. Kurtosis, on the other hand, appears to attain its highest values around the edges of bright polarization gradient regions, similar to the skewness. Hence, we conclude that the kurtosis of the polarization gradient is not an ideal statistic for studying interstellar turbulence. 

Although the mean and standard deviation of the polarization gradient appear to be promising statistics, they are not perfect, as a foreground depolarizing object will affect the values of the mean and standard deviation around the edges of the object. Topological statistics such as the genus may not be affected by foreground depolarizers, and hence may be suitable candidates for study, as suggested by \cite{Burkhart2012}.

\section{Discussion}
\label{discuss}

From the mosaics of $|\nabla{\bf{P}}|$ that we have produced, we find that there is significant gradient signal over approximately 70\% of the area of the survey. Based on our knowledge of the polarization horizon throughout the survey, these gradients are typically caused by magneto-ionic material within $3$-$4$ kpc. In contrast, polarized intensity above the noise level is detected in virtually every direction in the data of \citet{Landecker2010}. This is partly a result of sensitivity, but there are definitely directions in which the polarized signal is smooth: turbulence is not seen in all directions. A prime example is the Fan region, ${120^{\circ}}<{\ell}<{160^{\circ}}$ for latitudes above ${b}\approx{1^{\circ}}$. In Figure~\ref{pgrad_cgps_all1} most of this area shows the hexagonal noise pattern and no significant $|\nabla{\bf{P}}|$ signal. This is a region of strong polarized intensity \citep{Landecker2010}, and the polarized signal here is very smooth, at least that part of it that lies within the polarization horizon. 

A contrast is found in the area ${70^{\circ}}<{\ell}<{85^{\circ}}$, around the perimeter of the Cygnus-X region. The gradient signal is low here simply because the polarized intensity is very low; again, the hexagonal noise pattern becomes prominent.

We observe that the polarization gradient structures vary with angular resolution. This suggests that the turbulent features that are revealed by the polarization gradient method exist at different depths within the observed volume. The observations sample everything in a cone between the telescope and the polarization horizon. If the mechanisms that create turbulence are similar throughout the cone then the turbulent features will have approximately the same physical size throughout and convolution will sample structure at different distances. For example, the polarization gradient structures observed at poor angular resolution may be closer to the observer, or be very large features that are far away. The polarization gradient structures also are not correlated with the diffuse Stokes I over most of the survey, implying that the polarization gradient filaments are mostly caused by Faraday rotation and depolarization (either within the emitting region, or between the source and the observer), rather than changes in the strength of the magnetic field perpendicular to our line of sight, within the emitting region.  

In the northern latitude extension of the CGPS, we find that the polarization gradient features are stronger at the disk-halo transition, located at $b \approx 8^{\circ}$, than in the mid-plane of the Galaxy, or at higher Galactic latitudes. This could imply that the turbulence in the disk-halo transition region is very strong, and this can perhaps be ascribed to infall processes (\citealp{Putman2012}, \citealp{Fraternali2013}) creating MHD instabilities.  Our sample of the disk-halo transition covers only a small range of longitudes, and it is important that more extensive observations be made to verify our suggestion.

We have found that there appears to be a morphological difference in the polarization gradients observed towards the anti-center and inner Galaxy, with the gradients observed in the inner Galaxy appearing to be smaller and less elongated. This may indicate that turbulence in the inner Galaxy is stronger on small scales, or that the observed turbulence is further away than the turbulence observed towards the anti-center, for example in the ``window'' of the polarization horizon that lies between $\ell \approx 60^{\circ}$ and $\ell \approx 80^{\circ}$.

Although we observe morphological differences between the polarization gradients observed towards the anti-center and inner Galaxy, we find that the skewness of the polarization gradient does not depend on Galactic longitude, for any evaluation box size. Additionally, we find that despite there being numerous ``double-jump'' features throughout the CGPS, denoting shocks, the skewness rarely attains values indicative of supersonic turbulence coincident with these features. If the skewness were a reliable probe of turbulence, then this would indicate that the regime of turbulence does not change very much throughout the observed region of the Galaxy. However, as the skewness of the polarization gradient is very sensitive to the angular resolution of the observation, and the size of the evaluation box, we believe that the skewness primarily probes the number and magnitude of inhomogeneities within the evaluation box. If a suitably uniform area of the sky can be found, then the skewness may probe turbulence in this area, although high angular resolution observations will be required to properly probe the turbulence.  

Our conclusion that the skewness of the polarization gradient is an unreliable statistic affects our interpretation of the results presented by \cite{Gaensler2011}, \citet{Iacobelli2014} and \cite{Sun2014}. \cite{Gaensler2011} introduced the polarization gradient method, and suggested that the magneto-ionic turbulence observed in the SGPS test region is transonic, based on the morphology of the polarization gradients they observed and a measured skewness of $0.3$. \cite{Iacobelli2014} applied the gradient technique to the entire Southern sky, at an angular resolution of $10.8'$. Their work is complementary to ours in the sense that we concentrate mainly on emission in the Galactic plane while they avoid the plane and concentrate their analysis on higher latitudes. Unfortunately, there is no overlap between our survey region and theirs. \citet{Iacobelli2014} analyze PDF statistics in eleven regions of size $25^{\circ} \times 25^{\circ}$. Out of these regions, they find that five of them are `Faraday thin' (little Faraday rotation occurring within the emission region), and deduce from the skewness of the polarization gradient that the turbulence is sub- to transonic in these areas. Two of the regions lie in the Galactic plane, and they report skewness values of $1.56$ and $1.70$ for these regions. \cite{Sun2014} analyzed S-PASS data for the Galactic plane within ${10^{\circ}}<{\ell}<{34^{\circ}}$ and $|b| < 5^{\circ}$, at a frequency of $2.3$ GHz with $10'$ resolution. From the skewness of the polarization gradients they observed, $1.9$, they concluded that the turbulence in the warm ionized medium in this direction of the sky was transonic. For their observations at $4.8$ GHz, \cite{Sun2014} found that the polarized structures were intrinsic to the emitting region, and so the skewness did not probe the turbulence in the warm-ionized medium. We notice that the skewness values reported by \cite{Iacobelli2014} and \cite{Sun2014} are similar, and that the skewness value reported by \cite{Gaensler2011} is similar to the skewness values we observe away from the edges of bright polarisation gradient regions. 

As the observations used by \cite{Iacobelli2014} and \cite{Sun2014} have similar angular resolution, and the resolution of the SGPS test region ($75''$) is similar to that of the CGPS, we conclude that the difference in measured skewness values may be caused by the different angular resolutions of these studies. This supports our finding that the skewness of the polarization gradient is more sensitive to the angular resolution and the size of the box used to calculate the skewness, than to the underlying turbulence in the warm-ionized medium. 

As the skewness of the polarization gradient is affected by foreground depolarizing regions, it is possible that the normalized polarization gradient, $|\nabla \textbf{P}|/|\textbf{P}|$ will not be affected by these regions. This may lead one to assume that the skewness of the normalized polarization gradient may be a suitable probe of turbulence. However, the normalized polarization gradient is noisy in depolarized regions, and so the measured skewness will likely be strongly affected by noise in the image. 

Whereas \cite{Iacobelli2014} and \cite{Sun2014} relied on the skewness of the polarization gradient to determine the regime of turbulence in their observed regions of the warm-ionized medium, \cite{Burkhart2012} used the genus statistic to show that the turbulence observed in the SGPS test region is transonic. There is still evidence that the regime of turbulence in the warm-ionized medium of the Milky Way is transonic. Future studies analyzing other statistics of the polarization gradient, or analyzing the statistics of the polarization gradient for different angles between the line of sight and the mean magnetic field, will be required to place reliable constraints on the regime of turbulence. In particular, the mean and standard deviation of the polarization gradient, or morphological statistics such as the genus may prove useful. New simulations of inhomogeneous turbulence also have the potential to shed light on how observed statistics of the polarization gradient should be interpreted.  

\section{Conclusions}
\label{concl}

We have presented polarization gradient data for the entire CGPS, ${\sim}1300$ square degrees imaged with arcminute resolution at $1.4$ GHz. We have found qualitative differences in the morphology of polarization gradient structures within the Galactic plane, and in the disk-halo transition region, which suggests that the regime of turbulence within the mid-plane of the Milky Way varies with position. To quantify changes in the regime of turbulence, we have calculated the skewness of the polarization gradient for the entire CGPS using a sliding box method. We found that the skewness of the polarization gradient acts as an edge-detector, as it attains its largest values on the edges of bright polarization gradient regions. Regions with moderate values of skewness may have an unknown contribution from such edges, meaning that the skewness of the polarization gradient does not directly probe turbulence. Furthermore, we find that the skewness maps observed are sensitive to the angular resolution and the size of the evaluation box used to calculate the skewness. We find no significant variation in skewness with longitude. 

These findings imply that the skewness of the polarization gradient does not probe the underlying turbulence, casting doubt on previous deductions of the regime of turbulence from the skewness of the polarization gradient. We do not believe that our findings are a result of how the CGPS data have been processed, as we have masked point sources, ensured that the smoothing procedure does not introduce false gradients around masks, and observe the same edge detection in the skewness map of the polarization gradient image of the Southern Galactic Plane Survey test region.

We conclude that the skewness and kurtosis of the polarization gradient are not ideal probes of the magneto-ionic turbulence revealed by polarization gradients as they are too sensitive to observed inhomogeneities, and so there is less evidence that turbulence in the warm-ionized medium is transonic. The mean and standard deviation of the polarization gradient, or morphological statistics of the gradient, may provide useful constraints on the regime of turbulence.

\section*{Acknowledgements}

The Canadian Galactic Plane Survey is a Canadian project with international partners, supported by the Natural Sciences and Engineering Research Council. We thank Blakesley Burkhart and Mordecai-Mark Mac Low for useful conversations, Sean Dougherty for his encouragement through the process of completing this paper, and the referee for insightful comments that have improved this paper. C.~A.~H. acknowledges financial support received via an Australian Postgraduate Award, and a Vice Chancellor's Research Scholarship awarded by the University of Sydney. B.~M.~G. acknowledges the support of the Natural Sciences and Engineering Research Council of Canada (NSERC) through grant RGPIN-2015-05948. N.~M.~M.-G. acknowledges the support of the Australian Research Council through grant FT150100024. E.~P. acknowledges the support from the European Research Council under the European Union's Seventh Framework Programme (FP/2007-2013) / ERC Grant Agreement n. 617199. The Dunlap Institute for Astronomy and Astrophysics is funded through an endowment established by the David Dunlap family and the University of Toronto. This research made use of APLpy, an open-source plotting package for Python hosted at http://aplpy.github.com.

\bibliography{ref_list}
\end{document}